\documentclass[11p]{article}
\usepackage[dvips]{graphicx}
\usepackage{amsmath,amssymb,amsfonts}
\begin{document}
\title{\bf Horava-Lifshitz early universe phase transition beyond detailed balance}
\author{F. Kheyri\thanks{%
email: f\_kheyri@sbu.ac.ir}\,\,, M. Khodadi\thanks{%
email: m.khodadi@sbu.ac.ir}\,\, and H. R. Sepangi\thanks{%
email: hr-sepangi@sbu.ac.ir}\\
{\small Department of Physics, Shahid Beheshti University, Evin,
Tehran 19839, Iran}}
\maketitle
\begin{abstract}
The early universe is believed to have undergone a QCD phase transition to hadrons at about $10\mu s$ after the big bang. We study such a
transition in the context of the non-detailed balance Horava-Lifshitz theory by investigating the effects of the dynamical coupling constant
$\lambda$ in a flat universe. The evolution of the relevant physical quantities, namely the energy density $\rho$, temperature $T$, scale factor
$a$ and  the Hubble parameter $H$ is investigated before, during and after the phase transition, assumed to be of first order. Also, in view of
the recent lattice QCD simulations data, we study a cross-over phase transition of the early universe whose results are  based on two different
sets of lattice data.
\end{abstract}

\section{Introduction}
Recently, a  power-counting, ultra-violet $(UV)$ renormalizable quantum
gravity theory has been proposed \cite{JHEP03 020 , Horava. Rev. D 79},
stirring a considerable amount of interest in its wake. This theory is
principally based on a scalar field theory introduced by Lifshitz \cite{Lifshitz}
to explain quantum critical phenomenon in condensed matter physics. In this theory
 the action is scale-invariant under the transformations
\begin{eqnarray}
t\rightarrow\ell^{z}t,\quad x^{i}\rightarrow \ell x^{i},
\end{eqnarray}
with $z = 2$, where $z$ represents the dynamical critical exponent.  However,
the theory proposed by Horava has the scaling dimension $z=3$ and is often
referred as  Horava-Lifshitz (HL) gravity and is fundamentally non-relativistic.
This theory does not have the full diffeomorphism invariance of general relativity,
admitting only  a local Galilean invariance. Einstein general relativity
may then be considered as emerging as an infra-red (IR) fixed point.
Invariably, such a theory may have interesting ramifications when applied to
cosmology and, not surprisingly, a great amount of interest has been generated
in this regard. In particular, study of the early universe in the context of the HL
gravity \cite{Calcagni, Kiritsis} would represent an interesting area of research.
A number of works have appeared over the past few years studying the early
universe in this context. In \cite{Kiritsis} it was suggested that the
divergence of the speed of light in the UV region may resolve the horizon
problem.
The phenomenological implications of the HL theory were addressed in \cite{13}.
Black hole solutions and related issues were studied very actively in \cite{14, 15, 16, 17}, some of the theoretical issues and generalizations of the HL model were discussed in \cite{20, 21, 22, 23, 24}. Needless to say, a myriad of other interesting topics and issues can be found in the literature.
In view of the above, the present work is devoted to the study of QCD phase transition in the context of HL gravity in the early universe.

According to standard cosmology, as the early universe expanded and cooled, it underwent a series of symmetry-breaking phase transitions where topological defects may have formed. During its early evolution, the universe is believed to have undergone  at least two phase transitions. The electroweak theory predicts that at about $100GeV$ there was a transition from a symmetric high temperature phase with massless gauge bosons to the Higgs phase, where the $SU(2)\times U(1)$ gauge symmetry is spontaneously broken and all the masses are generated.
The second transition, predicted by quantum chromodynamics (QCD), is believed to have occurred at a temperature of about $200MeV$ where there was a transition from a quark-gluon plasma to a plasma of light hadrons. At about the same energy we expect that the global chiral symmetry of QCD with massless fermions is spontaneously broken by the formation of a quark pair condensates. The nature of the QCD transition affects our understanding of the evolution of the universe, see \cite{Dominik J} for further insight. In a strong first-order phase transition the quark-gluon plasma supercools before bubbles of hadron gas are formed. These bubbles grow, collide and merge, during which gravitational waves could be produced \cite{Edward Witten}.
 Baryon-enriched nuggets could remain between the bubbles, contributing to dark matter. The hadronic phase is the initial condition for nucleosynthesis, so inhomogeneities in this phase could have a strong effect on its development \cite{J. H. Applegate and C. J. Hogan}.

 The nature of the phase transition involved in the early stages of the evolution of the universe has been a source of some debate. A first order phase transition was the basic assumption under which calculations had been done in the past. However, detailed lattice QCD calculations performed over the years has now upset the balance in favor of a smooth crossover phase transition.   It is therefore the purpose of the present work to study the formation of hadrons in the early stages
of the evolution of the universe based on the assumption that a smooth crossover phase transition has been responsible. For the sake of clarity and comparison, we have also performed our calculations based on the assumption of a first order phase transition which we will discuss in due course.

Fundamental to the study of any phase transition in the early universe is a knowledge of the equation of state (EoS) which plays an important role in the hot, strongly interacting matter  \cite{O.Philipsen, M.Laine} and hydrodynamic description of heavy ion collisions \cite{P. F. Kolb}. There are a number of models where the  equation of state in a first order phase transition is used for describing the collective flow in heavy ion collisions. However, as was generally mentioned above, lattice QCD suggests that the transition to the de-confined phase is only a crossover one \cite{Y. Aoki}. The effects of the EoS on all aspects of the behavior of the system under study is therefore of great importance. For example, it is not obvious to what extent the collective flow is sensitive to details of the equation of state. It is now well known that in an ideal fluid dynamics scenario, the anisotropy of the proton flow is most sensitive to the QCD equation of state \cite{Huovinen}.

Many attempts have been made in the last two decades to calculate the EoS on a lattice, see  \cite{P. Petreczky, C. DeTar}.
Lattice calculations are usually  done at finite temperatures and at fixed temporal extent $N_{\tau}$ where the temperature is changed by varying the lattice spacing $a$, $T=1/N_{\tau}a$ . The  thermodynamic observables are calculated through the calculation of the trace of the energy momentum tensor $(\epsilon-3p)$, known as  trace anomaly, which can be done in the high temperature region rather accurately which is not the case in the low temperature region. Some authors have studied an analysis of the trace anomaly within the Hadron Resonance Gas model (HRG) \cite{P. Huovinen}. Such studies show that to reproduce the lattice results for the asqtad and $p4$ actions of the hotQCD collaboration, it is mandatory  to distort the resonance spectrum away from the physical one, necessary  for taking into account the larger quark masses used in these lattice calculations.
So, to construct a realistic equation of state one could use the lattice data. In a number of works lattice calculations show a reasonably good agreement with  hadron resonance gas (HRG) model predictions at low temperature \cite{F.Karsch, S.Ejiri, M.Chengetal} and in  other works such distortions are not needed to describe the data \cite{JHEP 1009}.

The starting point when studying the evolution of the early universe, or in general any cosmological setting, are the Friedmann equations. These equations in HL gravity differ from those of the standard $4D$ cosmology in the presence of the parameter $\lambda$ and of the dark radiation term. We expect this deviation from the standard $4D$ cosmology to have noticeable effects on the cosmological phase transitions. The first-order phase transition have been studied in HL gravity \cite{Malihe}, in the context of brane-world  scenarios \cite{Davis} and in a Randall-Sundrum (RS) brane-world model \cite{T. harko}. It would therefore be interesting to study crossover phase transition by using the EoS resulting from  lattice studies. In what follows, we shall concentrate on the occurrence of  phase transition in the early universe in the context of  HL gravity, using the type of phase transition and EoS suggested by lattice studies, that is, assuming a smooth crossover quark-hadron phase transition. We also present a comparison between these results and those when the phase transition is considered to be of first order.
\section{ Horava-Lifshitz gravity}
 Horava theory is constructed on the basic assumption of anisotropic scaling between space and time, i.e.
\begin{eqnarray}
t\rightarrow\ell^{3}t,\quad x^{i}\rightarrow\ell x.
\end{eqnarray}
In a theory with anisotropic scaling, time and space are fundamentally distinct. It is then natural to use the ADM formalism for which the metric is given by
\begin{eqnarray}
ds^2=-N^2 dt^{2}+g_{ij}\left(dx^{i}+N^{i}dt\right)\left(dx^{j}+N^{j}dt\right),
\end{eqnarray}
where $g_{ij}$, $N$ and $ N_{i}$ are the spatial metric, lapse and shift functions, respectively. These variables are dynamical variables which under the above scaling, scale as
 \begin{eqnarray}
 N\rightarrow N,\quad g_{ij}\rightarrow g_{ij},\quad N_{i}\rightarrow\ell^{2}N_{i},\quad N^{i}\rightarrow\ell^{-2}N^{i}.
 \end{eqnarray}
 The action of HL theory is decomposed into a kinetic, a potential and the matter part as
\begin{eqnarray}
S_{HL}=S_{k}+S_{v}+S_{m}.
\end{eqnarray}
The kinetic action in the Horava-Lifshitz gravity (HL) is naturally given as
\begin{eqnarray}\label{e2}
S_{k}=\int dtdx^{3}N\sqrt{g}\frac{2}{\kappa^{2}}\left(K_{ij}K^{ij}-\lambda K^{2}\right),
\end{eqnarray}
where $K_{ij}$ $(K=K^{i}\,_{i})$ is the extrinsic curvature of spatial slices, defined by
\begin{eqnarray}
K_{ij}=\frac{1}{2N}(\partial_{t}  g_{ij}-\nabla_{i} N_{j}-\nabla_{j}N_{i}),
\end{eqnarray}
with $\nabla$ being the covariant derivative on the spatial slice whose metric $g_{ij}$ is used to raise and lower indices.
The most general potential for $z=3$ can be written as
\begin{eqnarray}
S_{v}&=&\frac{-2}{\kappa^2}\int dt d^{3}x \sqrt{g} N[\sigma+\gamma R+\gamma_{1}R^{2}+\gamma_{2}R_{ij}R^{ij}+\xi \varepsilon^{ijk}R_{il}\nabla_{j}R^{l}_{k}+\sigma_{1}R^{3}+\sigma_{2}RR^{ij}R_{ij}\\ \nonumber
&+&\sigma_{3}R^{j}_{i}R^{k}_{j}R^{i}_{k}
+\sigma_{4}\nabla_{i}R\nabla^{i}R+\sigma_{5}\nabla_{i}R_{jk}\nabla^{i}R^{jk}].
\end{eqnarray}
On the other hand, the above ten coefficients can be reduced to three, denoted by $\mu, \Lambda_{W}, \zeta$, using the detailed balance condition \cite{Horava. Rev. D 79, JHEP03 020 }
\begin{eqnarray} \label{e1}
&\sigma=-\frac{3\kappa^{4}\mu^{2} \Lambda_{W}^{2}}{16(3\lambda-1)}, \gamma=\frac{\kappa^{4}\mu^{2}\Lambda_{W}}{16(3\lambda-1)}, \gamma_{1}=\frac{\kappa^{4}\mu^{2}(1-4\lambda)}{64(3\lambda-1)},\gamma_{2}=\frac{\kappa^{4}\mu^{2}}{16},\xi=-\frac{\kappa^{4}\mu}{4\zeta^{2}},\\ \nonumber
&\sigma_{1}=\frac{\kappa^{4}}{8\zeta^{4}},\sigma_{2}=-\frac{5\kappa^{4}}{8\zeta^{4}},\sigma_{3}=\frac{3\kappa^{4}}{4\zeta^{4}},\sigma_{4}=-\frac{3\kappa^{4}}{32\zeta^{4}},\sigma_{5}=\frac{\kappa^{4}}{4\zeta^{4}},
\end{eqnarray}
where $\kappa^{2}[\equiv\frac{8\pi G}{c^{4}}]$ and $\Lambda_{W}$ are the Einstein coupling and cosmological constants, respectively, and finally
\begin{eqnarray}
 S_{m}=\int dtdx^{3}N\sqrt{g}L_{m},
\end{eqnarray}
where $L_{m}=L_{m}(N,N_{i},g_{ij},\phi)$ is the lagrangian density of matter fields, denoted collectively by $\phi$.
In the IR limit the HL action simplifies to
\begin{eqnarray}\hspace{-0.5cm}
S\rightarrow S_{I}\simeq \int dt dx^{3}N\sqrt{g}[\alpha(K_{ij}K^{ij}-\lambda K^{2})+\gamma R+\sigma].
\end{eqnarray}
Defining $x^{0}=ct$ and
\begin{eqnarray}\hspace{-0.3cm}
\lambda=1,\quad c=\sqrt{\frac{\gamma}{\alpha}},\quad 16\pi G=\sqrt{\frac{\gamma}{\alpha^{3}}},\quad \Lambda _{EH}=-\frac{\sigma}{2\alpha},
\end{eqnarray}
will reduce this action to that of the Einstein-Hilbert
\begin{eqnarray}
S_{EH}&=&\frac{1}{16\pi G}\int dx^{4}\sqrt{\widetilde{g}}(\widetilde{R}_{4}-2\Lambda_{EH})\\ \nonumber
&=&\frac{1}{16\pi G}\int dt dx^{3} \sqrt{g} N[K_{ij}K^{ij}-K^{2}+(R-2\Lambda)].
\end{eqnarray}
The full space-time metric $\widetilde{g}_{\mu\nu}$ is given by
\begin{eqnarray}
\widetilde{g}_{00}=-N^{2}+g^{ij}N_{i}N_{j},\quad \widetilde{g}_{0i}=N_{i},\nonumber\\
\widetilde{g}_{ij}=g_{ij},\quad \sqrt{\widetilde{g}}=N\sqrt{g}.\hspace{1.5cm}
\end{eqnarray}
There is a condition inferred from the study of some cosmological models \cite{A. Wang}, given by
\begin{eqnarray}
\alpha(3\lambda-1)>0.
\end{eqnarray}

\section{Field equations in Horava-Lifshitz gravity}
Now, in order to focus on a cosmological framework, we use the $FRW$ metric
\begin{eqnarray}\label{e4}
N=1,\quad g_{ij}=a^{2}(t)\gamma_{ij},\quad N^{i}=0,
\end{eqnarray}
with
\begin{eqnarray}\label{e5}
\gamma_{ij}dx^{i}dx^{j}=\frac{dr^{2}}{1-kr^{2}}+r^{2}d\Omega_{2}^{2}.
\end{eqnarray}
The homogeneous and isotropic metric is then written as
\begin{eqnarray}
ds^{2}=-dt^{2}+a(t)^{2}\left[\frac{dr^{2}}{1-kr^{2}}+r^{2}(d\theta^{2}+\sin^{2}\theta\, d\phi^{2})\right],
\end{eqnarray}
where $k=-1,\,0,\,1$ represents an open, flat or closed universe, respectively. For the matter energy-momentum tensor we restrict our analysis to the case of a perfect fluid given in comoving coordinates by
\begin{eqnarray}
T_{\mu\nu}=(\rho+p)u_{\mu}u_{\nu}+pg_{\mu\nu},\quad u_{\mu}=-\delta^{t}_{\mu},
\end{eqnarray}
where $\rho$, $p$ and $u_{\mu}$ are the energy density, isotropic pressure and four-velocity of the cosmological fluid.
We obtain the gravitational field equations of the HL theory, corresponding to the line element (\ref{e4}) and (\ref{e5}) as
\begin{eqnarray}\label{e10}
3(3\lambda-1)H^{2}=\frac{\kappa^{2}}{2}\rho+6\left[\frac{\sigma}{6}+\frac{k\gamma}{a^{2}}+
\frac{2k^{2}(3\gamma_{1}+\gamma_{2})}{a^{4}}+\frac{4k^{3}(9\sigma_{1}+3\sigma_{2}+\sigma_{3})}{a^{6}}\right],
\end{eqnarray}
\begin{eqnarray}
(3\lambda-1)(\dot{H}+\frac{3}{2}H^{2})=-\frac{\kappa^{2}}{4}p-3\left[-\frac{\sigma}{6}-\frac{k\gamma}{3a^{2}}+
\frac{2k^{3}(9\sigma_{1}+3\sigma_{2}+\sigma_{3})}{a^{6}}\right],
\end{eqnarray}
with
\begin{eqnarray}\label{e11}
\dot{\rho}+3H(\rho+p)=0.
\end{eqnarray}
Since the early universe is known to be flat to within a good approximation, in the follow we shall focus on the non detailed balance condition with $k=0$.
\section{First order quark-hadron phase transition}
In this section, we present a discussion on the derivation of thermodynamical quantities of the quark-hadron phase transition which corresponds to the first order phase transition. It is now generally accepted that such phase transition has occurred at the cosmological QCD scale about $t\sim 10^{-5}s$ after the big bang and at a temperature $T\sim200MeV$ where the Hubble radius was about $10$ km. Note that the mass inside the Hubble volume is about $1M_{\bigodot}$.

In order to study the first order quark-hadron phase transition it is necessary to specify the equation of state of the matter in both the quark and hadron states. So, at this stage, let us at first discuss the EoS for a first order phase transition \cite{T. harko} which in the quark phase can generally be given in the form
\begin{eqnarray}\label{e18}
\rho_{q}=3a_{q}T^{4}+V(T),\quad p_{q}=a_{q}T^{4}-V(T),
\end{eqnarray}
where $a_{q}=(\pi^{2}/90)g_{q}$, with $g_{q}=16+(21/2)N_{F}+14.25=51.25$ and $N_{F}=2$. V(T) is a self-interacting potential, which is expressed as \cite{Borghini}
\begin{eqnarray}
V(T)=B+\gamma_{T}T^{2}-\alpha_{T}T^{4},
\end{eqnarray}
where $B$ is the bag pressure constant and, according to the results obtained in low energy hadron spectroscopy, heavy ion collisions and phenomenological fits of light hadron properties, the constants are given as $B^{1/4}\in(100-200)MeV$ \cite{D. Lee}, $\alpha_{T}=7\pi^{2}/20$ and $\gamma_{T}=m_{s}^{2}/4$ with $m_{s}$ being the mass of the strange quark in the range $m_{s}\in(60-200)MeV$. The shape of the potential $V$ reflects the physical model in which the quark fields are interacting with a chiral field formed out of the $\pi$ meson and a scalar field.
For the hadron phase, we choose a cosmological fluid  consisting of an ideal gas of massless pions and nucleons described by the Maxwell-Boltzmann statistics, having an energy density $\rho_{h}$ and pressure $p_{h}$, respectively. The equation of state is given by
\begin{eqnarray}\label{e19}
p_{h}(t)=\frac{1}{3}\rho_{h}(T)=a_{\pi}T^{4},
\end{eqnarray}
where $a_{\pi}=(\pi^{2}/90)g_{h}$ and $g_{h}=17.25$.
The critical temperature $T_{c}$ is defined by the condition $p_{q}(T_{c})=p_{h}(T_{c})$ \cite{Kajantieh}, and is given by
\begin{eqnarray}
T_{c}=\sqrt{\frac{\gamma_{T}+\sqrt{\gamma_{T}^{2}+4(a_{q}+\alpha_{T}-a_{\pi})B}}{2(a_{q}+\alpha_{T}-a_{\pi})}}.
\end{eqnarray}
For $m_{s}=200MeV$ and $B^{1/4}=200MeV$ the transition temperature is of the order of $T_{c}\approx125 MeV$. A first order phase transition incorporates discontinuities for all the physical quantities like the energy density, pressure and entropy across the critical curve. The quark and hadron energy densities ratio at the critical temperature, $\rho_{q}(T_{c})/\rho_{h}(T_{c})$, is of the order $3.62$ for $m_{s}=200MeV$ and $B^{1/4}=200MeV$.
\section{Dynamics of the universe during the first order transition}
\begin{figure}
\includegraphics[width=8cm]{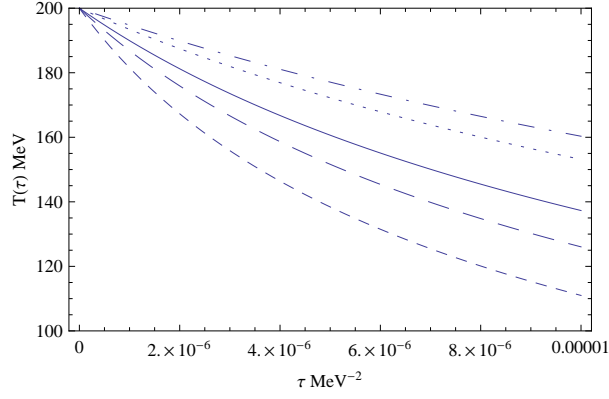}
\centering\caption{\footnotesize  Variation of  temperature in the $T>T_{c}$ region as a function of $\tau=\kappa t$ in a first order phase transition for different values of $ \lambda: \lambda=3$ (dotted-dashed curve), $\lambda =2$ (dotted curve), $\lambda =1$ (solid curve), $\lambda =0.7$ (long dashed curve), and $ \lambda =0.5$ (short dashed curve), we set $\frac{\sigma}{\kappa^{2}}=10^{6}$ .}{\label{fig.q} \small }
\end{figure}

\begin{figure}
\includegraphics[width=8cm]{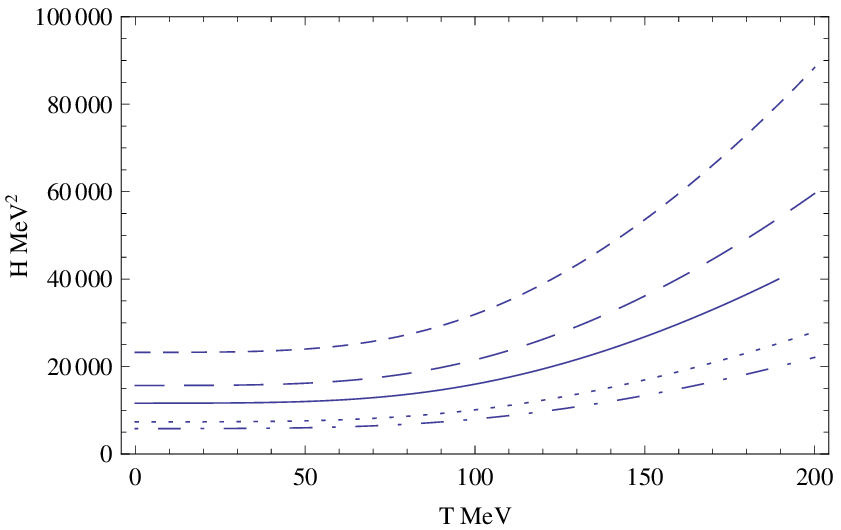}
\centering\caption{\footnotesize  Variation of the Habble parameter in the $T>T_{c}$ region as a function of temperature in a first order phase transition for different values of $\lambda: \lambda=3$ (dotted-dashed curve), $\lambda =2$ (dotted curve), $\lambda =1$ (solid curve), $\lambda =0.7$ (long dashed curve), and $ \lambda =0.5$ (short dashed curve), we set $\frac{\sigma}{\kappa^{2}}=10^{6}$ .}{\label{fig.H q} \small }
\end{figure}

\begin{figure}
\includegraphics[width=8cm]{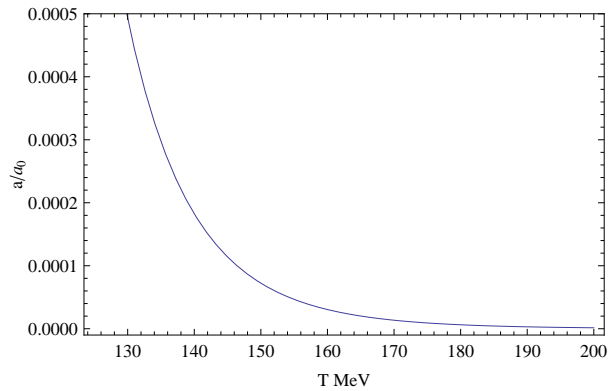}
\centering\caption{\footnotesize  Variation of the scale factor in the $T>T_{c}$ region as a function of temperature in a first order phase transition for different values of $ \lambda: \lambda=3$ (dotted-dashed curve), $\lambda =2$ (dotted curve), $\lambda =1$ (solid curve), $\lambda =0.7$ (long dashed curve), and $ \lambda =0.5$ (short dashed curve), we set $\frac{\sigma}{\kappa^{2}}=10^{6}$ .}{\label{fig.a q} \small }
\end{figure}

\begin{figure}
\includegraphics[width=8cm]{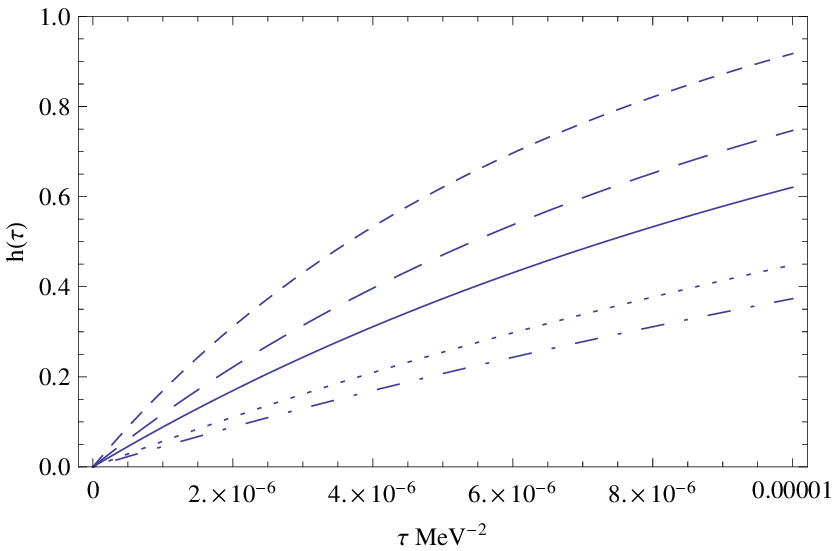}
\centering\caption{\footnotesize  Variation of the hadron fraction for $T=T_{c}$ as a function of $\tau=\kappa t$ in a first order phase transition for different values of $ \lambda: \lambda=3$ (dotted-dashed curve), $\lambda =2$ (dotted curve), $\lambda =1$ (solid curve), $\lambda =0.7$ (long dashed curve), and $ \lambda =0.5$ (short dashed curve), we set $\frac{\sigma}{\kappa^{2}}=10^{6}$ .}{\label{f. mix} \small }
\end{figure}

\begin{figure}
\includegraphics[width=8cm]{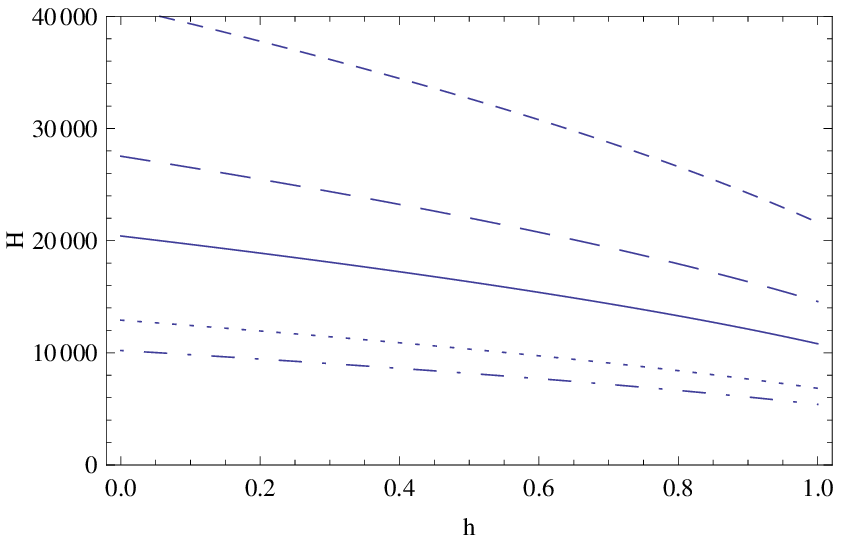}
\centering\caption{\footnotesize  Variation of the Habble parameter in $T=T_{c}$ case as a function of $hadron fraction$ in a first order phase transition for different values of $ \lambda: \lambda=3$ (dotted-dashed curve), $\lambda =2$ (dotted curve), $\lambda =1$ (solid curve), $\lambda =0.7$ (long dashed curve), and $ \lambda =0.5$ (short dashed curve), we set $\frac{\sigma}{\kappa^{2}}=10^{6}$.}{\label{fig.H mix} \small }
\end{figure}

\begin{figure}
\includegraphics[width=8cm]{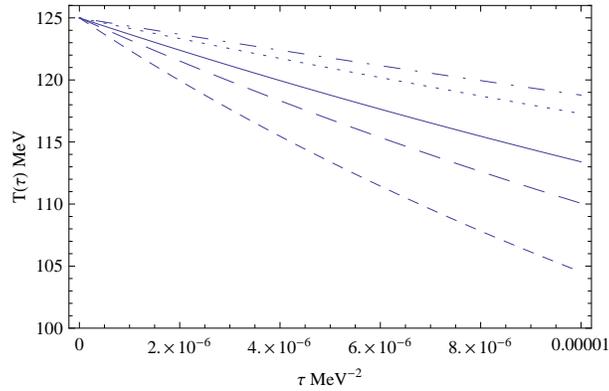}
\centering\caption{\footnotesize  Variation of  temperature for $T<T_{c}$ as a function of $\tau=\kappa t$ in a first order phase transition for different values of $ \lambda: \lambda=3$ (dotted-dashed curve), $\lambda =2$ (dotted curve), $\lambda =1$ (solid curve), $\lambda =0.7$ (long dashed curve), and $ \lambda =0.5$ (short dashed curve), we set $\frac{\sigma}{\kappa^{2}}=10^{6}$.}{\label{fig.hadron} \small }
\end{figure}

\begin{figure}
\includegraphics[width=8cm]{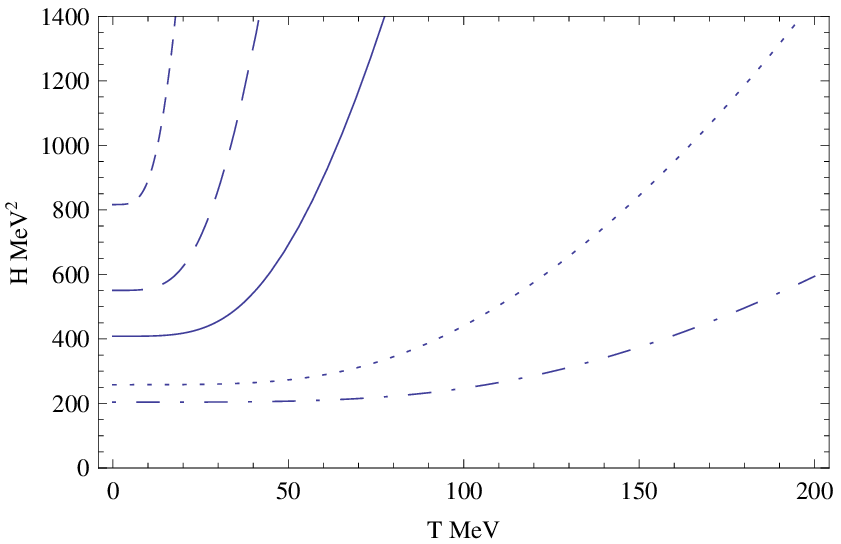}
\centering\caption{\footnotesize  Variation of the Habble parameter for $T<T_{c}$ as a function of temperature in a first order phase transition for different values of $\lambda:\lambda=3$ (dotted-dashed curve), $\lambda =2$ (dotted curve), $\lambda =1$ (solid curve), $\lambda =0.7$ (long dashed curve), and $ \lambda =0.5$ (short dashed curve), we set $\frac{\sigma}{\kappa^{2}}=10^{6}$.}{\label{fig.H hadron} \small }
\end{figure}

After determining the equation of state, the physical quantities like the energy density $\rho$, temperature $T$,  scale factor $a(t)$ and Habble parameter $H$ can be monitored during the phase transition. These quantities are determined by the Eqs.\,(\ref{e10}), (\ref{e11}), (\ref{e18}) and (\ref{e19}). Naturally, there are three stages to distinguish; before the phase transition where $T>T_{c}$ and the universe is likely to be in the quark phase, during the phase transition where $T=T_{c}$ and all the quark matter is changed to hadrons and finally, after the phase transition where $T<T_{c}$ and the universe is in the hadron phase. Now, using the equations of state of the quark matter, equation\,(\ref{e11}), the variation of cosmological quantities such as the Hubble parameter, scale factor and temperature can be written as
\begin{eqnarray}\label{H q}
H(T)=\frac{\dot{a}}{a}=-\frac{3a_{q}-\alpha_{T}}{3a_{q}}\frac{\dot{T}}{T}-\frac{1}{6}\frac{\gamma_{T}}{a_{q}}\frac{\dot{T}}{T^{3}},
\end{eqnarray}
where after integration we have
\begin{eqnarray}\label{a q}
a(T)=\frac{a_{0}}{3a_{q}}T^{(\alpha_{T}-3a_{q})} \exp\left(\frac{1}{12}\frac{\gamma_{T}}{a_{q}}\frac{1}{T^{2}}\right),
\end{eqnarray}
where $a_{0}$ is a constant of the integration and the variation of temperature for $T>T_{c}$ can be written as
\begin{eqnarray}\label{e20}
\frac{dT}{d\tau}=-\frac{T^{3}}{(1-\frac{\alpha_{T}}{3a_{q}})T^{2}+\frac{\gamma_{T}}{6a_{q}}}
\times \sqrt{\frac{1}{6(3\lambda-1)}\left[(3a_{q}-\alpha_{T})T^{4}+\gamma_{T}T^{2}+B
+2\sigma\right]}
\end{eqnarray}
Figures  \ref{fig.q}, \ref{fig.H q} and \ref{fig.a q}  show the behavior of temperature as a function of the cosmic time $\tau$ , Hubble parameter and scale factor as a function of temperature in a HL world filled with quark matter for different values of $\lambda$ .
For $T = T_{c}$, the temperature and  pressure are constant. In this case quantities like the entropy $ S = sa^{3}$ and  enthalpy $W =(\rho + p)a^{3}$ are conserved. For later convenience, we replace $\rho(t)$ by $h(t)$, so that the volume fraction of matter in the hadron phase is given by
\begin{eqnarray}\label{h f}
\rho(t)=\rho_{h}(t)+\rho_{q}(1-h(t))=\rho_{Q}(1+mh(t)),
\end{eqnarray}
where $m=\frac{\rho_{H}}{\rho_{Q}}-1$. The beginning of the phase transition is characterized by $h(t_{c})= 0$ where $t_{c}$ is the time scale and $\rho(t_{c}) \equiv\rho_{Q}$, while the end of the transition is characterized by $h(t_{h})=1$ with $t_{h}$ being the time signaling the end and corresponding to $\rho(t_{h})\equiv\rho_{H}$.
To find the variation of hadron fraction we substitute equation\,(\ref{h f}) in the field equations and obtain
\begin{eqnarray}
\frac{dh}{dt}=-\frac{3(1+rh(t))}{r}\left[\frac{\kappa^{2}\rho_{Q}}{6(3\lambda-1)}(1+mh(t))+\frac{\sigma}{3(3\lambda-1)}\right]^{1/2},
\end{eqnarray}
where $r=(\rho_{H}-\rho_{Q})/(\rho_{Q}+p_{c})$. Figure \ref{f. mix} shows variation of the hadron fraction $h(\tau)$ as a function of $\tau$ for different values of $\lambda$.
One may present  variation of the Hubble parameter in this case as a function of the hadron fraction by using the field equation and equation\, (\ref{h f}), for which the result is shown in figure \ref{fig.H mix} for different values of $\lambda$.
For $t>t_{h}$ the universe enters into the hadronic phase. For the phase transition temperature of $T_{c}=125MeV$ we have $\rho_{Q}=5\times10^{9}MeV^{4}$ and $\rho_{H}=1.38\times10^{9}MeV^{4}$, respectively. For the same value of the temperature the value of the pressure of the cosmological fluid during the phase transition is $p_{c}=4.6\times10^{8}MeV^{4}$. Following the same procedure as that in the quark phase, that is,  substituting equation\,(\ref{e19}) into (\ref{e11}), the evolution of the scale factor can be obtained
where $a(\tau_{h})$ is now determined by using the conservation equation (\ref{e11}) and equations of state for $T<T_{c}$ which leads to  $a(T)=a(\tau_{h})T_{c}/T$.
In the last stage of the first order phase transition, variation of temperature in the hadron phase ($T<T_{c}$) for a flat universe leads to
\begin{eqnarray}
\frac{dT}{dt}=-T\sqrt{\frac{\kappa^{2}a_{\pi}T^{4}}{2(3\lambda-1)}+\frac{\sigma}{3(3\lambda-1)}}.
\end{eqnarray}
The variation of temperature in a first order phase transition as a function of $\tau=kt$ and also variation of the Hubble parameter as a function of temperature are represented for different values of the coupling constant $\lambda$, in figures \,\ref{fig.hadron} and \ref{fig.H hadron}, respectively.
This parameter shows that small values of the coupling constant $\lambda$ reduce the temperature in these regions. Also, from figures \,\ref{fig.q} and \ref{fig.hadron} it is seen that for $\lambda>1$, the temperature is higher than in general relativity ($\lambda=1$) whereas for $1/3<\lambda<1$ the temperature is much lower.
In the next section, we shall consider the evolution of temperature for a crossover phase transitions in Horava-Lifshitz theory.
\section{Crossover quark-hadron phase transition}
Indeed, our work is based on results obtained from two different sets of  lattice calculations.
The hot QCD collaboration analysis is based on  two different improved staggered fermion actions, asqtad and $p4$, with a physical strange quark mass and physical $u$ and $d$ quark masses, $ m_{s}/m_{u,d}= 10$. It shows that the transition region lies in the range
$T =(185-195)MeV$, for recent published results  see \cite{A. Bazavov}. Results presented by the other collaborations Points to the value of the transition temperature in the range $150-170 MeV$, and  changes with the observable used to define it \cite{Aoki 2, Aoki 3}. The most recent results for several physical quantities are presented in \cite{JHEP 1009} with the equation of state at $N_{t}=6,8,10$ and $12$ given in  \cite{borsanyi}.
We shall rely on the above two cases of analysis to determine the EoS in the context of the crossover phase transition in this work. In the first case, the equation of states in both phases are given by the hadron resonance gas (HRG) model at low temperatures and lattice QCD data at high temperatures, which are combined with a simple parametrization given in \cite{P. Huovinen}.  The equation of state has been calculated with the so-called $p4$ and asqtad actions on lattices with temporal extent $N_{\tau} = 4, 6$ and $8$ \cite{C. Bernard, M. Cheng, A. Bazavov}.

Calculation of the thermodynamical observables in lattice QCD proceeds through the calculation of the trace of the energy momentum
tensor $\Theta(T)=\epsilon-3p$, also known as the trace anomaly. Different thermodynamical observables can be obtained from the trace anomaly through
integration. The pressure can be written as
\begin{eqnarray}\label{e22}
\frac{p(T)}{T^{4}}-\frac{p(T_{0})}{T_{0}^{4}}=\int_{T_{0}}^{T}\frac{dT'}{T'^{5}}(\rho-3p).
\end{eqnarray}
Here $T_{0}$ is an arbitrary temperature value that is usually chosen in the low-temperature regime where the pressure is exponentially small.
This method is known as the integral method \cite{G. boyd}.
Finite temperature lattice calculations are mostly performed at a fixed  temporal extent $N_{\tau}$
with the temperature being changed by varying the lattice spacing a, ($T =1/N_{\tau}a)$.
Therefore, the lattice spacing becomes smaller as the temperature is increased. As a consequence, the trace anomaly can be calculated in the high temperature region accurately, whereas at low temperature is affected by  the large discretization effects. Thus, to construct a more realistic equation of state one should use the lattice data for the trace anomaly in the high temperature region, $T > 250MeV$, and  the HRG model otherwise, $T<180MeV$.

The HRG model with modified masses is believed to describe the lattice data rather well up to temperatures of about $180MeV$. In the middle region, $180MeV<T<250MeV$, the HRG model cannot be trusted, whereas discretization effects in lattice calculations could be large. Therefore, at high temperatures the trace anomaly can be parameterized in a polynomial form in general, see e.g. \cite{A. Bazavov}. We shall thus use the following ansatz for the high temperature region
\begin{eqnarray}
\frac{(\rho-3p)}{T^{4}}=\frac{d_{2}}{T^{2}}+\frac{d_{4}}{T^{4}}+\frac{c_{1}}{T^{n_{1}}}+\frac{c_{2}}{T^{n_{2}}}.
\end{eqnarray}
This form is shown to work well in the temperature range of interest and  is flexible enough to be matched to the HRG result in the low temperature region. The resulting parametrization of the trace anomaly and  QCD equation of state obtained using these requirements are labeled as $s95p-v1$, $s95n-v1$
and $s90f-v1$, as shown in table 1. The labels ``s95'' and ``s90'' refer to the fraction of the ideal entropy density reached at $T=800MeV$ (95\% and
90\%, respectively). The labels $p$, $n$ and $f$ refer to a specific treatment of the peak of the trace anomaly or its matching to the HRG. Values of the parameters $T_{0}$, $d_{2}$, $d_{4}$, $c_{1}$, $c_{2}$, $n_{1}$ and $n_{2}$ in  are  given in table 1. The following form shows a parameterized form for the HRG 
\begin{eqnarray}
\frac{(\rho-3p)}{T^{4}}=a_{1}T+a_{2}T^{3}+a_{3}T^{4}+a_{4}T^{10},
\end{eqnarray}
where $a_{1}=4.654GeV^{-1}, a_{2}=-879GeV^{-3}, a_{3}=8081GeV^{-4}$ and $a_{4}=-7039000GeV^{-10}$.
The importance of the above parametrization and integral method is that we can obtain the thermodynamical quantities for the low temperature region as
\begin{eqnarray}\label{e15}
p(T)=(\alpha_{0}+\beta_{0})T^{4}+a_{1}T^{5}+\frac{1}{3}a_{2}T^{7}+\frac{1}{4}a_{3}T^{8}
+\frac{1}{10}a_{4}T^{14},
\end{eqnarray}
\begin{eqnarray}\label{e12}
\rho(T)=3(\alpha_{0}+\beta_{0})T^{4}+4a_{1}T^{5}+2a_{2}T^{7}+\frac{7}{4}a_{3}T^{8}
+\frac{13}{10}a_{4}T^{14},
\end{eqnarray}
where $\alpha_{0}$ and $\beta_{0}$ are the value of integration in the lower limit and the value of $p_{0}^{4}/T_{0}^{4}$ when $T_{0}=70 MeV$, respectively.
For the high temperature region, the pressure and energy density are given by
\begin{eqnarray}\label{e13}
p(T)=-\frac{1}{4}d_{4}-\frac{d_{2}}{2}T^{2}+(\alpha_{i}+\beta_{i})T^{4}-
\frac{c_{1}}{n_{1}}T^{4-n_{1}}-\frac{c_{2}}{n_{2}}{T^{4-n_{2}}},
\end{eqnarray}
\begin{eqnarray}\label{e14}
\rho(T)=3(\alpha_{i}+\beta_{i})T^{4}-\frac{1}{2}d_{2}T^{2}+
\frac{1}{4}d_{4}+\frac{c_{1}}{n_{1}}(n_{1}-3)T^{4-n_{1}}
+\frac{c_{2}}{n_{2}}(n_{2}-3)T^{4-n_{2}},
\end{eqnarray}
where $i=1,2,3$ corresponds to different parameterizations which are given in table $1$. So, $\alpha_{i}$ and $\beta_{i}$ refer to the values of integration in the lower limit and the value of $p_{0}/T_{0}^{4}$ in each case. Equations of state are also available in a tabulated form in \cite{online}.

As was mentioned above, there is another set of data which we will use for determining the EoS in the present work for which a simple parametrization of the trace anomaly, given for $n_{f}=2+1$ flavor  and also for an estimate of $n_{f}=2+1+1$ flavor equation of state, can be found in \cite{borsanyi}. We notice that these results are obtained for Symanzik improved gauge and a stout-link improved staggered fermion action on lattices with temporal extent $N_{\tau} = 6, 8$ and $10$. The thermodynamical quantities are presented as functions of  temperature in the range  $100\cdots1000MeV$. A global parametrization of the trace anomaly as a function of temperature is given by the following fitted function
\begin{eqnarray}\label{e25}
\frac{\rho-3p}{T^{4}}=\exp\left(-h_{1}/t-h_{2}/t^{2}\right)
\times&\left[h_{0}+\frac{f_{0}(\tanh(f_{1}t+f_{2})+1)}{1+g_{1}t+g_{2}t^{2}}\right],
\end{eqnarray}
where the dimensionless $t$ variable is defined as $t=T/(200MeV)$. The parameters can be found in 2able $2$. In this case, the numerical solution is used for determining the EoS. The results of this calculation are presented in the following section.

\begin{table}
\caption{\footnotesize The values of parameters for different fits to the trace anomaly \cite{P. Huovinen}.}
\begin{tabular}{|c|c|c|c| c| c| c| c|}
  \hline
  & $d_{2}(GeV)$ &$d_{4}(GeV^{4})$ &$c_{1}(GeV^{n_{1}})$&$c_{2}(GeV^{n_{2}})$&$n_{1}$&$n_{2}$&$T_{0}(MeV)$\\ \hline
  $s95p$ &$0.2660$ &$2.403\times10^{-3} $&$-2.809\times10^{-7} $&$6.073\times10^{-23}$&$ 10$&$ 30$&$183.8$\\
$s95n$&$0.2654$&$6.563\times10^{-3}$&$-4.370\times10^{-5}$&$5.774\times10^{-6}$&$8$&$9$&$171.8$\\
$s90f$&$0.2495$&$1.355\times10^{-2}$&$-3.237\times10^{-3}$&$1.439\times10^{-14}$&$5$&$18$&$170.0$\\
  \hline
\end{tabular}
\end{table}

\begin{table}
\caption{\footnotesize The values of  parameters appearing  in equation \,(\ref{e25}) describing the trace anomaly for $n_{f}=2+1$ and  $n_{f}=2+1+1$ \cite{borsanyi}.}
\begin{tabular}{|c|c|c|c|c|c|c|c|c|}
  \hline
   $n_{f}$ &$h_{0}$ &$h_{1}$&$h_{2}$&$f_{0}$&$f_{1}$&$f_{2}$&$g_{1}$&$g_{2}$\\ \hline
  $2+1$ &$0.1396$ &$-0.1800 $&$0.0350$&$2.76$&$6.79$&$ -5.29$&$-0.47$&$1.04$\\ \hline
  $2+1+1$&$0.1396$&$-0.1800$&$0.0350$&$5.59$&$7.34$&$-5.60$&$1.42$&$0.50$\\ \hline
\end{tabular}
\end{table}

\section{Dynamical evolution in the crossover phase transition}
In this section we shall study the physical quantities of interest, that is, energy density $\rho$ and temperature $T$. For the first set of lattice data \cite{A. Bazavov} mentioned before,  these quantities are obtained from the Friedmann equation\,(\ref{e10}), conservation equation\,(\ref{e11}) and the equations of state determined by equations\,(\ref{e15}) and \,(\ref{e12}) for low temperature region and equations\,(\ref{e13}) and \,(\ref{e14}) for high temperature region in the crossover phase transition. By assuming a first order phase transition, these quantities can be obtained from the Friedmann equation\,(\ref{e10}), conservation equation\,(\ref{e11}) and equations of state (\ref{e18}) and (\ref{e19}), determined for $T>T_{c}$ and $T<T_{c}$, respectively.
Using the conservation equation \,(\ref{e11}) and equation of state which is used for describing the high temperature region, we have
\begin{eqnarray}\label{e16}
\frac{\dot{a}}{a}&=& \\
&-&\frac{12(\alpha_{i}+\beta_{i})T^{3}-d_{2}T+\frac{c_{1}}{n_{1}}(n_{1}-3)(4-n_{1})T^{3-n_{1}}+\frac{c_{2}}{n_{2}}
(n_{2}-3)(4-n_{2})T^{3-n_{2}}}{3(-d_{2}T^{2}+4(\alpha_{i}+\beta_{i})T^{4}+\frac{c_{1}}{n_{1}}(n_{1}-4)T^{4-n_{1}}+\frac{c_{2}}{n_{2}}(n_{2}-4)
T^{4-n_{2}})}\dot{T}.\nonumber
\end{eqnarray}
Substituting equation\,(\ref{e16}) in equation\,(\ref{e10}), the basic equation describing the evolution of temperature of the universe in the
high temperature region can be written as
\begin{align}\label{e17}
\frac{dT}{dt}&=-\frac{3(-d_{2}T^{2}+4(\alpha_{i}+\beta_{i})T^{4}+\frac{c_{1}}{n_{1}}(n_{1}-4)T^{4-n_{1}}+\frac{c_{2}}{n_{2}}(n_{2}-4)
T^{4-n_{2}})}{12(\alpha_{i}+\beta_{i})T^{3}-d_{2}T+\frac{c_{1}}{n_{1}}(n_{1}-3)(4-n_{1})T^{3-n_{1}}+\frac{c_{2}}{n_{2}}
(n_{2}-3)(4-n_{2})T^{3-n_{2}}} \nonumber\\
&\times\bigg[\kappa^2\frac{3(\alpha_{i}+\beta_{i})T^{4}-\frac{1}{2}d_{2}T^{2}+
\frac{1}{4}d_{4}+\frac{c_{1}}{n_{1}}(n_{1}-3)T^{4-n_{1}}
+\frac{c_{2}}{n_{2}}(n_{2}-3)T^{4-n_{2}}}{6(3\lambda-1)}\nonumber\\
&+\frac{\sigma}{3(3\lambda-1)}+\frac{k A}{a^{2}}
+\frac{k^{2}B}{a^{4}(t)}+\frac{k^{3}C}{a^{6}(t)}\bigg]^\frac{1}{2},
\end{align}
where
\begin{eqnarray}
A=\frac{2\gamma}{(3\lambda-1)},\,
B=\frac{4(3\gamma_{1}+\gamma_{2})}{(3\lambda-1)},\,
C=\frac{8(9\sigma_{1}+3\sigma_{2}+\sigma_{3})}{(3\lambda-1)},
\end{eqnarray}
and other parameters are determined as in the previous section. The coefficients of the curvature are absent in a spatially flat universe $(k=0)$, thus the coefficients $A$, $B$ and $C$ are  zero, so that equation\,(\ref{e17}) reduces to
\begin{eqnarray}
\frac{dT}{dt}&=&-\frac{3(-d_{2}T^{2}+4(\alpha_{i}+\beta_{i})T^{4}+\frac{c_{1}}{n_{1}}(n_{1}-4)T^{4-n_{1}}+\frac{c_{2}}{n_{2}}(n_{2}-4)
T^{4-n_{2}})}{12(\alpha_{i}+\beta_{i})T^{3}-d_{2}T+\frac{c_{1}}{n_{1}}(n_{1}-3)(4-n_{1})T^{3-n_{1}}+\frac{c_{2}}{n_{2}}
(n_{2}-3)(4-n_{2})T^{3-n_{2}}} \nonumber\\
&\times&\left[\frac{\kappa^2(3(\alpha_{i}+\beta_{i})T^{4}-\frac{1}{2}d_{2}T^{2}+
\frac{1}{4}d_{4}+\frac{c_{1}}{n_{1}}(n_{1}-3)T^{4-n_{1}}
+\frac{c_{2}}{n_{2}}(n_{2}-3)T^{4-n_{2}})+2\sigma}{6(3\lambda-1)}
\right]^\frac{1}{2},
\end{eqnarray}
which has been modified relative to the general relativity result with parameter $\lambda$.

\begin{figure}
\includegraphics[width=8cm]{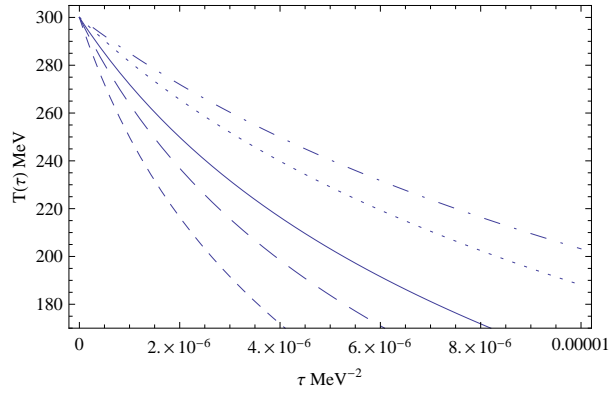}
\centering\caption{\footnotesize  Variation of temperature in the high temperature region as a function of $\tau=\kappa t$ for the first parametrization corresponding to $T_{0}=170.0 MeV$. The behavior of temperature is shown for different values of $ \lambda: \lambda=3$ (dotted-dashed curve), $\lambda =2$ (dotted curve), $\lambda =1$ (solid curve), $\lambda =0.7$ (long dashed curve) and $ \lambda =0.5$ (short dashed curve), we set $\frac{\sigma}{\kappa^{2}}=10^{10}$.}{\label{fig.1} \small}
\end{figure}

Moving forward, we first study the first order phase transition where the curvature term is zero. We then focus attention on the study of a crossover phase transition with the same conditions. The variation of  temperature as a function of  $\tau=\kappa t$ in the high temperature region is presented for a flat universe for different values of the coupling constant $\lambda$, in figure \,\ref{fig.1}. The behavior of this parameter shows that at small values of the coupling constant $\lambda$, the temperature is reduced in this era. As can be seen in these figures, for $\lambda>1$ the temperature is higher than the general relativity case where $\lambda=1$, whereas for $1/3<\lambda<1$ the temperature is much lower.

\begin{figure}
\includegraphics[width=8cm]{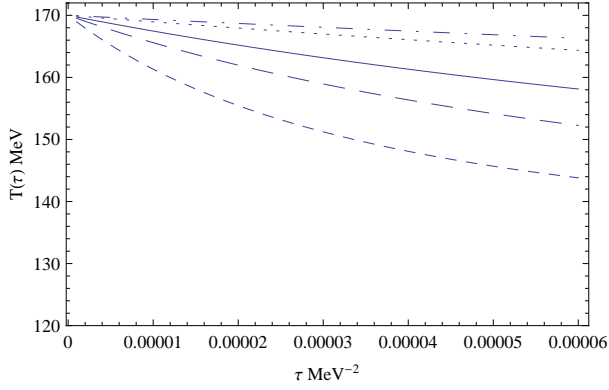}\caption{Variation of  temperature of the low temperature region as a function of $\tau=\kappa t$ for different values of $ \lambda: \lambda=3$ (dotted-dashed curve), $\lambda =2$ (dotted curve), $\lambda =1$ (solid curve), $\lambda =0.7$ (long dashed curve), and $ \lambda =0.5$ (short dashed curve), we set $\frac{\sigma}{\kappa^{2}}=10^{6}$.}{\label{fig.2} \small}
\end{figure}

For the low temperature region, $T<170 MeV$,  using the conservation equation\,(\ref{e11}) and  equations of state  (\ref{e13}) and (\ref{e14}) and also equation\,(\ref{e10}), the basic equation describing the evolution of temperature of the universe  can be written as
\begin{align}
\frac{dT}{dt}&=-\frac{12(\alpha_{0}+\beta_{0})T+15a_{1}T^{2}+7a_{2}T^{4}+6a_{3}T^{5}+
\frac{21}{5}a_{4}T^{11}}{12(\alpha_{0}+\beta_{0})+20a_{1}T
+14a_{2}T^{3}+14a_{3}T^{4}+\frac{91}{5}a_{4}T^{10}} \nonumber \\
&\times\left[\kappa^{2}\frac{3(\alpha_{0}+\beta_{0})T^{4}+4a_{1}T^{5}+2a_{2}T^{7}+\frac{7}{4}a_{3}T^{8}+
\frac{13}{10}a_{4}T^{14}}{6(3\lambda-1)}
 +\frac{\sigma}{(3\lambda-1)}\right]^{\frac{1}{2}},\hspace{6cm}
\end{align}
where a dot represents derivative respect to time. The variation of temperature as a function of  parameter $\tau=\kappa t$ in the low temperature region is represented, in the case of  $k=0$, for different values of the coupling constant $\lambda$, in figure \ref{fig.2}. The behavior of
this parameter shows that  small values of the coupling constant $\lambda$ reduce the effective temperature. Figure \,\ref{fig.2} shows that for $\lambda>1$, the temperature is higher than in general relativity ($\lambda=1$) whereas for $1/3<\lambda<1$ the temperature is much lower, again relative to general relativity.

\begin{figure}
\includegraphics[width=8cm]{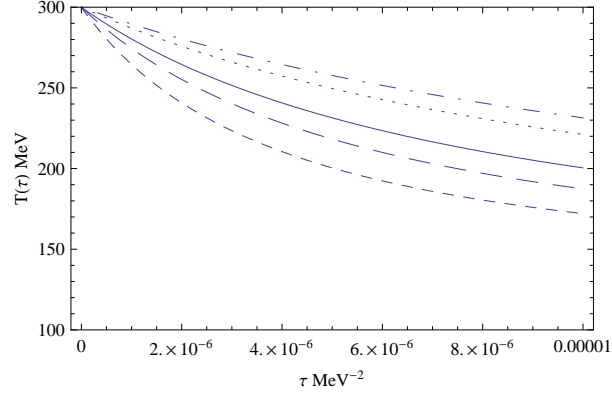}
\centering\caption{\footnotesize  Variation of  temperature for $n_{f}=2+1$ flavor as a function of $\tau=\kappa t$ in a smooth cross-over phase transition  for different values of $ \lambda: \lambda=3$ (dotted-dashed curve), $\lambda =2$ (dotted curve), $\lambda =1$ (solid curve), $\lambda =0.7$ (long dashed curve), and $ \lambda =0.5$ (short dashed curve) where we set $\frac{\sigma}{\kappa^{2}}=10^{6}$.}{\label{fig.7} \small}
\end{figure}

\begin{figure}
\includegraphics[width=8cm]{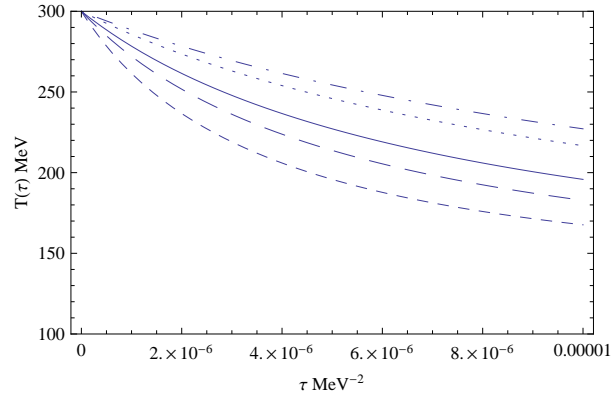}
\centering\caption{\footnotesize  Variation of temperature for $n_{f}=2+1+1$ flavor as a function of $\tau=\kappa t$ in a smooth cross-over phase transition  for different values of $ \lambda: \lambda=3$ (dotted-dashed curve), $\lambda =2$ (dotted curve), $\lambda =1$ (solid curve), $\lambda =0.7$ (long dashed curve), and $ \lambda =0.5$ (short dashed curve) where we set $\frac{\sigma}{\kappa^{2}}=10^{6}$.}{\label{fig.8} \small}
\end{figure}

At this stage, we turn to the second set of lattice data \cite{borsanyi} mentioned before and repeat the above calculations. However, this time there is no analytical expression for the equation of state and one has to resort to numerical calculations to obtain the evolution of temperature for the parameterization given in equation\ref{e25}.
Variation of  temperature for $n_{f}=2+1$ and $n_{f}=2+1+1$ flavors for different values of $ \lambda$ is presented in Figs. \ref{fig.7} and \ref{fig.8}, respectively. One realizes that the effect of $ \lambda$ on the variation of temperature in the early universe in this case is the same as
the results obtained by the ``hot QCD'' collaboration.
\section{Conclusions and Remarks}
In this work, we have studied the QCD phase transition in the context of Horava-Lifshitz gravity in the early universe, where differences from what is predicted by general relativity theory are likely to be noticeable. We investigated the effects of the coupling constant on the evolution of physical quantities, relevant to the physical description of the early universe in the case of first order and smooth crossover phase transitions. We used phenomenological equation of state which led to first order phase transition and investigated the influence of the coupling constant on the variation of temperature and Hubble parameter. The results show that the Hubble parameter decreases with time and that an increasing coupling constant decrease the value of this parameter before and after the phase transition. In addition, higher values of the coupling constant increase the effective temperature of the quark and hadron phases and also decrease the hadron fraction in mixed phase. In this work, two different sets of lattice simulation data which led to two different cases for the equation of state with smooth crossover phase transition, are used. In the first case, it led to the energy density and  temperature in both high ($T>250MeV$) and low ($T<170MeV$) temperature regions with the equation of state obtained by the ``hot QCD'' collaboration which is believed to be responsible for the emergence of hadrons from the quarks and gluons. In the second case, we used results obtained with the lattice data provided by the Wuppertal-Budabest collaboration. Comparison of these two different sets of results  shows  good agreement for the influence of the coupling constant on the variation of  temperature in the context of the HL gravity. We found that the temperature evolution of the early universe in Horava gravity is different from the idealized standard FRW model, so that for a small value of the coupling constant $\lambda$, phase transition occurs and results in describing the effective temperature in both high and low temperature regions. We also noted that the general behavior of the physical quantities mentioned above would not differ markedly if a first order phase transition is assumed to have occurred.



\begin{thebibliography}{9}
\bibitem{JHEP03 020 }P. Horava, JHEP  0903 (2009) 020 [arXiv:hep-th/0812.4267].
\bibitem{Horava. Rev. D 79}P. Horava, Phys. Rev. D 79 (2009) 084008 [arXiv:hep-th/0901.3775].
\bibitem{Lifshitz}E. M. Lifshitz, ZH. Eksp. Toer. Fiz. 11 (1941) 255-269.
\bibitem{Kiritsis} E. Kiritsis, G. Kofinas, Nucl. Phys. B 821 (2009) 467 [arXiv:hep-th/0904.1334].
\bibitem{Calcagni}G. Calcagni, JHEP 0909 (2009) 112 [arXiv:hep-th/0904.0829].
\bibitem{13} S. Mukohyama, K. Nakayama, F. Takahashi and S. Yokoyama, Phys. Lett. B 679 (2009) 6-9 [arXiv:hep-th/0905.0055].
\bibitem{14} H. Lu, J. Mei and C. N. Pope, Phys. Rev. Lett 103 (2009) 091301 [arXiv:hep-th/0904.1595].
\bibitem{15} Y. S. Myung and Y. W. Kim, Eur. phys. J. C 68 (2010) 265 [arXiv:hep-th/0905.0179].
\bibitem{16}A. Kehagias and K. Sfetsos, Phys. Lett. B 678 (2009) 123 [arXiv:hep-th/0905.0477].
\bibitem{17} R. G. Cai, L. M. Caoand and N. Ohta, Phys. Lett. B 679 (2009) 504-509 [arXiv:hep-th/0905.0751].
\bibitem{20} J. Kluson, JHEP 0907 (2009) 079 [arXiv:hep-th/0904.1343].
\bibitem{21}H. Nastase, TIT-HEP-596 (2009) 11pp [arXiv:hep-th/0904.3604].
\bibitem{22}R. G. Cai, Y. Liuand and Y. W. Sun, JHEP 0906 (2009) 010 [arXiv:hep-th/0904.4104].
\bibitem{23}T. Sotiriou, M. Visser and S. Weinfurtner, Phys. Rev. Lett. 102 (2009) 251601 [arXiv:hep-th/0904.4464].
\bibitem{24}D. Orlando and S. Reffert, Class. Quant. Grav. 26 (2009) 155021 [arXiv:hep-th/0905.0301].
\bibitem{Dominik J} D. J. Schwarz, Annalen Phys. 12 (2003) 220–270 [arXiv:astro-ph/0303574].
\bibitem{Edward Witten} E. Witten, Phys. Rev. D 30 (1984) 272.
\bibitem{J. H. Applegate and C. J. Hogan} J. H. Applegate and C. J. Hogan, Phys. Rev. D 31 (1985) 3037–3045.
\bibitem{O.Philipsen} M. Hindmarsh and O. Philipsen, Phys. Rev. D 71 (2005) 087302 [arXiv:hep-th/0501232].
\bibitem{M.Laine} M. Laine, PoS LAT (2006) 014 [arXiv:hep-lat/0612023].
\bibitem{P. F. Kolb}
     P.F. Kolb and U. W. Heinz [arXiv:nucl-th/0305084];
     P. Huovinen [arXiv:nucl-th/0305064].
\bibitem{Y. Aoki} Y. Aoki, G. Endrodi, Z. Fodor, S. D. Katz and K. K. Szabo, Nature 443 (2006) 675 [arXiv:hep-lat/0611014].
\bibitem{Huovinen}P. Huovinen, Nucl. Phys. A 761 (2005) 296 [arXiv:nucl-th/0505036].
\bibitem{P. Petreczky} P. Petreczky, Nucl. Phys. B (Proc. Suppl.) 140 (2005) 78 [arXiv:hep-lat/04009139].
\bibitem{C. DeTar} C. DeTar, PoS LAT (2008) 001 [arXiv:hep-lat/0811.2429].
\bibitem{P. Huovinen} P. Huovinen and P. Petreczky, Nucl. Phys. A 837 (2010) 26-53 [Arxiv:hep-ph/0912.2541].
\bibitem{F.Karsch} F. Karsch, K. Redlichand and A. Tawfi, Phys. Lett .B 571 (2003) 67 [arXiv:hep-ph/0306208].
\bibitem {S.Ejiri} S. Ejiri, F. Karschand and K. Redlich, Phys. Lett. B 633 (2006) 275 [arXiv:hep-ph/0509051].
\bibitem {M.Chengetal} M. Chengetal, Phys. Rev. D 79 (2009) 074505 [arXiv:hep-lat/0811.1006].
\bibitem{Malihe} M. Heydari-Fard, Gen Relativ Gravit. 42 (2010) 2729-2742.
\bibitem{Davis}S. C. Davis, W. B. Perkins, A. C. Davis and I. R. Vernon, Phys. Rev. D 63 (2001) 083518.
\bibitem{T. harko} G. De Risi, T. Harko, F.S.N. Lobo and C. S. J. Pun, Nucl. Phys. B 805 (2008) 190 [arXiv:gr-qc/0807.3066].
\bibitem{A. Wang} A. Wang and Y. Wu, JCAP 0907 (2009) 012.
\bibitem{C. Bernard} C. Bernard et al., Phys. Rev. D 75 (2007) 094505 [arXiv:hep-lat/0611031].
\bibitem{M. Cheng} M. Cheng et al., Phys. Rev. D 77 (2008) 014511 [arXiv:hep-lat/0710.0354].
\bibitem{online}https:$//wiki.bnl.gov/hhic/index.php/Lattice_{-}calculations_{-}of_{-}\\
Equation_of_State and https://wiki.bnl.gov/TECHQM/index.php/Bulk_Evolution$.
\bibitem{A. Bazavov} A. Bazavov et al., Phys. Rev. D 80 (2009) 014504, [arXiv:hep-lat/0903.4379].
\bibitem{G. boyd} G. Boyd, J. Engels, F. Karsch, E. Laermann, C. Legeland, M. Lutgemeier and B. Petersson, Nucl. Phys. B 469 (1996) 419 [arxiv:hep-lat/9602007].
\bibitem{Borghini} N. Borghini, W. N. Cottingham and R. Vinh Mau, J. Phys. G 26 (2000) 771 [arXiv:hep-ph/0001284].
\bibitem{D. Lee}T. D. Lee and Y. Pang, Phys. Rept. 221 (1992) 251.
\bibitem{Kajantieh} K. Kajantie and H. Kurki-Suonio, Phys. Rev. D 34 (1986) 1719.
\bibitem{JHEP 1009}S. Borsanyi et al., [Wuppertal-Budapest Collaboration], JHEP 1009 (2010) 073 [arXiv:hep-lat/1005.3508].
\bibitem{Aoki 2}Y. Aoki, Z. Fodor, S. D. Katz and K. K. Szabo, Phys. Lett. B 643, 46 (2006).
\bibitem{Aoki 3}Y. Aoki, S. Borsanyi, S. Durr, Z. Fodor, S. D. Katz, S. Krieg and K. K. Szabo, JHEP 0906, 088 (2009) [arXiv:hep-lat/0903.4155].
\bibitem{borsanyi} S. Borsanyi et al. JHEP 1011 (2010) 077 [arXiv:hep-lat/1007.2580].
\end{thebibliography}
\end{document}